\documentclass[pra,aps,reprint,superscriptaddress,amsmath]{revtex4-1}

\usepackage{bm}
\usepackage{graphicx}
\usepackage{dsfont}
\usepackage{xcolor}
\usepackage[utf8]{inputenc}

\newcommand{\bra}[1]{\langle #1 | \,}
\newcommand{\ket}[1]{\, | #1 \rangle}
\newcommand{\braket}[2]{\langle #1 | #2 \rangle}
\newcommand{\expv}[1]{\langle #1 \rangle}

\newcommand{\om}{\omega}

\newcommand{\la}{\lambda}
\newcommand{\eps}{\epsilon}
\newcommand{\veps}{\varepsilon}

\begin{document}

\title{Coherent interface between optical and microwave photons\\
on an integrated superconducting atom chip}

\author{David Petrosyan}
\affiliation{Institute of Electronic Structure and Laser,
Foundation for Research and Technology -- Hellas, GR-70013 Heraklion, Crete, Greece}
\affiliation{Center for Quantum Science, Physikalisches Institut, Eberhard Karls Universit\"at T\"ubingen, D-72076 T\"ubingen, Germany}

\author{J\'ozsef Fort\'agh}
\affiliation{Center for Quantum Science, Physikalisches Institut, Eberhard Karls Universit\"at T\"ubingen, D-72076 T\"ubingen, Germany}

\author{Gershon Kurizki}
\affiliation{Department of Chemical and Biological Physics, Weizmann Institute of Science, Rehovot 7610001, Israel}

\date{\today}

\begin{abstract}
Sub-wavelength arrays of atoms exhibit remarkable optical properties, analogous to those of phased array antennas,  
such as collimated directional emission or nearly perfect reflection of light near the collective resonance frequency.    
We propose to use a single-sheet sub-wavelength array of atoms as a switchable mirror 
to achieve a coherent interface between propagating optical photons and microwave photons 
in a superconducting coplanar waveguide resonator. 
In the proposed setup, the atomic array is located near the surface of the integrated superconducting chip 
containing the microwave cavity and optical waveguide. 
A driving laser couples the excited atomic state to Rydberg states with strong microwave transition.
Then the presence or absence of a microwave photon in the superconducting cavity makes the atomic array 
transparent or reflective to the incoming optical pulses of proper frequency and finite bandwidth.
\end{abstract}

\maketitle


\section{Introduction}

Systems  which hybridize modules with different quantum mechanical functionalities, 
alias hybrid quantum systems \cite{Xiang2013,Kurizki2015,Clerk2020}, have the potential to combine the merits of these functionalities 
and thereby become key to diverse quantum technologies, particularly in the domain of quantum information processing and communication.
Here we consider a system that hybridizes superconducting microwave and optical elements with cold trapped atoms 
to achieve an efficient interface between the microwave and optical photons.    
Coherent interfaces between microwave and optical radiation \cite{Lauk2020,Petrosyan2019,Covey2019,Tu2022,Kumar2023,Sahu2023} 
are required for interconnecting superconducting circuits, which are the best platform for quantum information processing 
operating in the microwave domain, via optical photons, which are the best carriers of quantum information over long distances
\cite{Kimble2008,Wehner2018}. 

The present scheme follows a pathway of great topical interest:
the interaction of light with regular arrays of strongly interacting atoms  
\cite{Bettles2016,Facchinetti2016,Asenjo-Garcia2017,Shahmoon2017,Solomons2023,Grankin2018,Guimond2019,Rui2020,Srakaew2022,Manzoni12018,Cardoner2021,Pedersen2022,Zhang2022}, governed by the cooperative Dicke effect \cite{Dicke1954} modified by the resonant dipole-dipole exchange interactions between the atoms \cite{Lehmberg1970,James1993,Thirunamachandran}.
Subwavelength arrays of atoms possess cooperative resonances corresponding to sub- or super-radiant optical modes and can serve
as, e.g., perfect optical mirrors \cite{Bettles2016,Shahmoon2017,Solomons2023,Guimond2019,Rui2020,Srakaew2022} or 
tailored, highly-efficient photon emitters into the desired spatial modes \cite{Asenjo-Garcia2017,Grankin2018}. 
Upon exciting the atoms to the strongly interacting Rydberg states by additional lasers, 
highly nonlinear interactions between single photons in such systems can also be achieved \cite{Cardoner2021,Zhang2022}.
Here, by combining the nearly perfect reflection of optical photons from a subwavelength atomic array,
electromagnetically induced transparency (EIT) \cite{EITrev2005} that randeres the array fully transparent, 
and strong coupling of Rydberg atomic transition to microwave radiation which switches on and off the EIT, 
we realize a coherent interface between the optical and microwave fields. 


\section{Array of two-level atoms interacting with a probe field}
\label{sec:formalism}

\begin{figure}[t]
\centering
\includegraphics[width=1.0\linewidth]{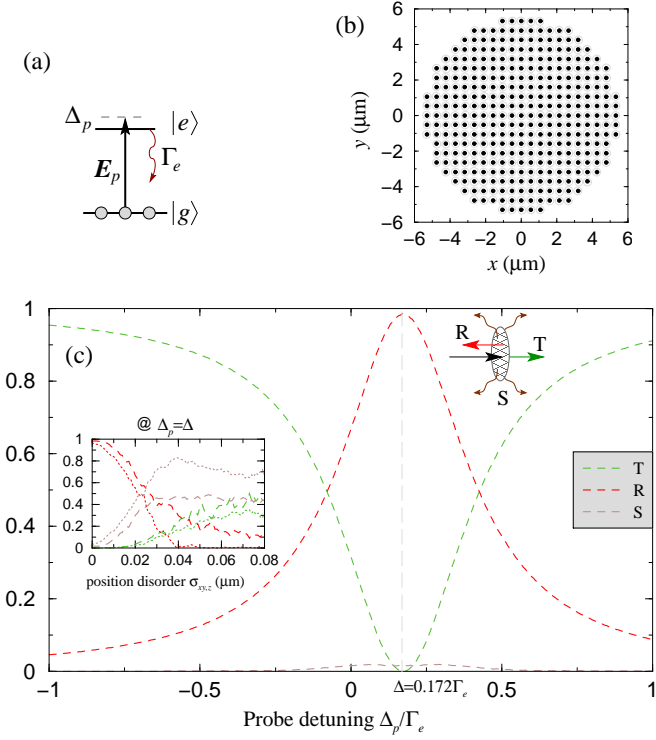}
\caption{(a) Level scheme of atoms interacting with the probe field $\bm{E}_p$ detuned by $\Delta_p$ from the atomic transition resonance, 
while the excited atomic state $\ket{e}$ decays to the ground state  $\ket{g}$ with rate $\Gamma_e = 2\pi \times 6\:$MHz.
(b) Two-dimensional array of atoms (black filled circles) with the lattice spacing $s=532\:$nm smaller than the wavelength $\lambda_e = 780\:$nm 
of the atomic transition $\ket{e} \to \ket{g}$. The atomic positions can deviate from the equilibrium lattice positions (open gray circles).  
(c) Transmission (T, green), reflection (R, red) and scattering (S, brown) spectra for an incoming coherent probe pulse ($\bar{n}_p = 1$)
of duration $\tau = 2\:\mu$s (Gaussian envelope) in a Gaussian spatial mode with waist $w_0 = 3 \lambda_e = 2.34\:\mu$m
focused at, and normal to, the atomic layer in the $xy$ plane.
Inset shows the transmission, reflection and scattering of the probe field at the collective resonance frequency $\Delta_p = \Delta$  
($\Delta = 0.172 \Gamma_e \simeq 2\pi \times 1\:$MHz) 
vs the (Gaussian) atomic position uncertainly (standard deviations) $\sigma_x = \sigma_y$ while $\sigma_z = 0$ (dashed lines),
and $\sigma_z$ while $\sigma_x = \sigma_y=0.01\:\mu$m (dotted lines). 
The graph are obtained from Monte Carlo simulations of Eqs.~(\ref{eqs:atamplsge}) for the dissipative dynamics of the stochastic atomic wavefunction
averaged over 100-200 independent trajectories, in conjunction with Eqs.~(\ref{eq:alphaTR}) and (\ref{eq:expvsigj}).}
\label{fig:scheme0}
\end{figure}

We first revisit the collective radiative properties of spatially-periodic arrays of atoms, taking into account 
the temporal and spatial profiles of an incident probe field near-resonant with an atomic transition.

\subsection{Dynamics of the atoms}

Consider an ensemble of $N$ atoms interacting with a probe field $\bm{E}_p(\bm{r},t)$ on the 
transition $\ket{g} \to \ket{e}$ with detuning $\Delta_p = \omega_p - \omega_{e}$, where $\omega_p$ is the 
carrier frequency of the probe and $\omega_{e}$ is the atomic transition frequency, see Fig. \ref{fig:scheme0}(a). 
The atoms are trapped in a two-dimensional lattice with a subwavelength period, $s < \lambda_e = 2\pi c/\omega_e$, 
see Fig. \ref{fig:scheme0}(b). Upon eliminating the free-space radiation modes interacting with the atoms using the
Born-Markov approximation \cite{Lehmberg1970,James1993,Thirunamachandran}, we obtain the master equation that governs the
dynamics of the density operator $\hat{\rho}$ of the atomic system,
\begin{equation}
\partial_t \hat{\rho} = -\frac{i}{\hbar} [H, \hat{\rho}] + \mathcal{L}[\hat{\rho}] ,
\end{equation}
where $H = H_{\mathrm{af}} + H_{\mathrm{RDDI}}$ is the Hermitian Hamiltonian of the system and 
$\mathcal{L}[\rho]$ is the Lindblad operator describing the collective atomic decay,
as detailed below. 

The atom-field interaction Hamiltonian is given by 
\begin{equation}
H_{\mathrm{af}} = \sum_j^N \{ \hbar \omega_e \hat{\sigma}_{ee}^{j} - [\hat{\sigma}_{eg}^{(j)} \bm{\wp}_{eg} \cdot \bm{E}_p(\bm{r}_j) e^{- i \omega_p t} + \mathrm{H. c.}] \},
\end{equation}
where $\hat{\sigma}_{\mu \nu}^{j} \equiv \ket{\mu}_j\bra{\nu}$ is the projection ($\mu = \nu$) or transition ($\mu \neq \nu$) operator for $j$th atom,
$\bm{\wp}_{eg}$ is the atomic transition dipole moment, and 
$\bm{E}_p(\bm{r}) = \hat{\bm{e}}_p \phi(\bm{r}) \hat{a}_p$ is the quantized probe field with polarization $\hat{\bm{e}}_p$
in the spatial mode $\phi(\bm{r})$ with the photon annihilation operator $\hat{a}_p$. 

Next, the resonant dipole-dipole exchange interaction between the atoms $i$ and $j$ at positions $\bm{r}_{i}$ and $\bm{r}_{j}$ is described by 
\begin{equation}
H_{\mathrm{RDDI}} = \hbar \sum_{j\neq i} V_{ji} \hat{\sigma}_{eg}^{(j)} \hat{\sigma}_{ge}^{(i)} 
\end{equation}
with 
\begin{eqnarray*}
V_{ji} &=& - \frac{k_e^2}{\hbar \eps_0} \bm{\wp}_{eg}^* \cdot \mathrm{Re} [\bm{G}( \bm{r}_{j},\bm{r}_{i}, k_e)] \cdot  \bm{\wp}_{eg} 
\nonumber \\
&=& \frac{3 \Gamma_e}{4} \left\{ \big[ 1 - (\hat{\bm{\wp}} \cdot \hat{\bm{r}}_{ij})^2 \big] \frac{\cos (k_e r_{ij})}{k_e r_{ij}} \right.
\nonumber \\ & & \left.
- \big[ 1 - 3 (\hat{\bm{\wp}} \cdot \hat{\bm{r}}_{ij})^2 \big]
\left[ \frac{\sin (k_e r_{ij})}{(k_e r_{ij})^2} + \frac{\cos (k_e r_{ij})}{(k_e r_{ij})^3} \right] \right\}  , \;\;\; 
\end{eqnarray*}
where $\bm{G}( \bm{r},\bm{r}', k_e)$ is the diadic Green's tensor for the free electromagnetic field \cite{Lehmberg1970,James1993},
$k_e =\omega_e/c$, $\hat{\bm{\wp}} \equiv \frac{\bm{\wp}_{eg}}{\wp_{eg}}$ is the unit vector in the direction of the atomic
dipole moment, $\hat{\bm{r}}_{ij} \equiv \frac{\bm{r}_{ij}}{r_{ij}}$ is the unit vector along the direction
of the relative position vector $\bm{r}_{ij} = \bm{r}_{i} -\bm{r}_{j}$ between atoms $i$ and $j$, $r_{ij} \equiv |\bm{r}_{ij}|$
is the distance between the atoms, and $\Gamma_e = \frac{1}{4 \pi \eps_0} \frac{4 k_e^3 |\wp_{eg}|^2}{3 \hbar}$ is the usual spontaneous 
decay rate of an atom in the excited state $\ket{e}$ \cite{ScullyZubary1997,PLDP2007}, 

Finally the Lindblad operator for the atomic decay is given by \cite{Lehmberg1970,James1993,Thirunamachandran}
\begin{equation}
\mathcal{L}[\hat{\rho}] = \frac{1}{2} \sum_{ij} \Gamma_{ji} [ 2 \hat{\sigma}_{ge}^{(j)} \hat{\rho} \hat{\sigma}_{eg}^{(i)} - 
\{ \hat{\sigma}_{eg}^{(j)} \hat{\sigma}_{ge}^{(i)} , \hat{\rho} \} ] , \label{eq:lindblad}
\end{equation}
where
\begin{eqnarray*}
\Gamma_{ji} &=& \frac{k_e^2}{\hbar \eps_0} \bm{\wp}_{eg}^* \cdot \mathrm{Im} [\bm{G}( \bm{r}_{j},\bm{r}_{i}, k_e)] \cdot  \bm{\wp}_{eg} 
\nonumber \\
&=& \frac{3 \Gamma_e}{2} \left\{ \big[ 1 - (\hat{\bm{\wp}} \cdot \hat{\bm{r}}_{ij})^2 \big] \frac{\sin (k_e r_{ij})}{k_e r_{ij}} \right.
\nonumber \\ & &\left.
+ \big[ 1 - 3 (\hat{\bm{\wp}} \cdot \hat{\bm{r}}_{ij})^2 \big]
\left[ \frac{\cos (k_e r_{ij})}{(k_e r_{ij})^2} - \frac{\sin (k_e r_{ij})}{(k_e r_{ij})^3} \right]  \right\}  . \;\;\; 
\end{eqnarray*}

\subsubsection*{Stochastic wavefunction approach}

The dissipative dynamics of the atomic system can equivalently be simulated using the quantum Monte Carlo stochastic 
wavefunction approach \cite{PLDP2007,QMCWF}, in which the wavefunction of the system $\ket{\Psi}$ evolves according to the
Schr\"odinger equation $\partial_t \ket{\Psi} = -\frac{i}{\hbar} \tilde{H} \ket{\Psi}$ with the effective Hamiltonian
\begin{equation}
\tilde{H} = H_{\mathrm{af}} + H_{\mathrm{RDDI}} - \frac{i}{2} \sum_{ij} \Gamma_{ji} \hat{\sigma}_{eg}^{(j)} \hat{\sigma}_{ge}^{(i)}  ,
\end{equation}
where the imaginary term is inherited from the last term of the Lindblad operator (\ref{eq:lindblad}). 
Since the effective Hamiltonian is non-Hermitian, the norm of the wavefunction 
$|| \Psi(t) || = \sqrt{\braket{\Psi(t)}{\Psi(t)}}$ is not preserved and 
the evolution is interrupted by quantum jumps affected by the first term of the Lindblad operator.
We can diagonalize the matrix $\Gamma_{ji} = \sum_l P^{\dagger}_{jl}  \Gamma_l P_{li}$ to obtain
the collective decay rates $\Gamma_l$, and rewrite the Lindblad operator in the diagonal form, 
\begin{equation}
\mathcal{L}[\hat{\rho}] = \frac{1}{2} \sum_{l} \Gamma_{l} [ 2 \hat{\Sigma}_{l} \hat{\rho} \hat{\Sigma}_{l}^{\dagger} - 
\{ \hat{\Sigma}_{l}^{\dagger} \hat{\Sigma}_{l} , \hat{\rho} \}] ,
\end{equation}
where the collective jump operators are $\hat{\Sigma}_{l} = \sum_j  P_{lj} \hat{\sigma}_{ge}^{(j)}$ \cite{Jones2017}.
The relative probabilities for quantum jumps $\ket{\Psi} \to \hat{\Sigma}_{l} \ket{\Psi}$ on the collective 
decay channels $l$ are then determined by $\bra{\Psi} \hat{\Sigma}_{l}^{\dagger} \hat{\Sigma}_{l} \ket{\Psi}$.
The normalized wavefunction of the system at any time is given by 
$\ket{\bar{\Psi}(t)} = \ket{\Psi(t)}/|| \Psi(t) ||$ and
the expectation value of any observable $\mathcal{O}$ of the system
is obtained by averaging over many, $M \gg 1$, independently simulated
trajectories, $\expv{\mathcal{O}} = \mathrm{Tr} [\hat{\rho} \mathcal{O}]
= \frac{1}{M} \sum_m^M \bra{\bar{\Psi}_m}  \mathcal{O} \ket{\bar{\Psi}_m}$,
while the density operator is given by
$\hat{\rho}(t) = \frac{1}{M} \sum_m^M \ket{\bar{\Psi}_m(t)}
\bra{\bar{\Psi}_m(t)}$.

\subsection{Dynamics of the field}

Consider now the field radiated by the atoms,  
\begin{eqnarray}
\bm{E}_{\mathrm{rad}}(\bm{r},t) &=& \frac{k_e^2}{\eps_0} \sum_j \bm{G}( \bm{r}_{j},\bm{r}_{i}, k_e) \cdot  \bm{\wp}_{eg}  
\hat{\sigma}^j_{ge}(t - |\bm{r} -\bm{r}_j|/c)
\nonumber \\ 
&=& \frac{\wp_{eg} k_e^2}{4\pi \eps_0}
\sum_j \frac{e^{-i \om_e (t-|\bm{r} -\bm{r}_j|/c)}}{|\bm{r} -\bm{r}_j|} \hat{\sigma}^j_{ge}(t - |\bm{r} -\bm{r}_j|/c)
\nonumber \\ & & \qquad \qquad \times
[\mathds{I} - \hat{\bm{r}}_j \otimes \hat{\bm{r}}_j] \cdot \hat{\bm{\wp}} ,
\end{eqnarray}
where $\mathds{I}$ is the unity tensor and $\hat{\bm{r}}_j \equiv \frac{\bm{r} - \bm{r}_j}{|\bm{r} - \bm{r}_j|}$.
In the far-field region, $|\bm{r} -\bm{r}_j| \simeq r - (\bm{r} \cdot \bm{r}_j)/r = r - \hat{\bm{r}} \cdot \bm{r}_j$, 
for each polarization component $\hat{\bm{e}}_{\bm{r},\sigma} \perp \bm{r}$ of the radiated field we have
\[
E_{\mathrm{rad}}^{(\sigma)}(\bm{r},t) = (\hat{\bm{e}}_{\bm{r},\sigma} \cdot \hat{\bm{\wp}})
\frac{\wp_{eg} k_e^2}{4\pi \eps_0}  \frac{e^{i (k_er - \om_e t)}}{r}
\sum_j  \hat{\sigma}^j_{ge} e^{-i \bm{k}_e \cdot \bm{r}_j} , 
\]
where $\bm{k}_e \equiv k_e \hat{\bm{r}}$.

In general, the total field in any position is given by the superposition of the incoming field 
and the field radiated by the atoms,
\begin{equation}
\bm{E}_{\mathrm{tot}}(\bm{r},t) = \bm{E}_{\mathrm{in}}(\bm{r},t) + \bm{E}_{\mathrm{rad}}(\bm{r},t).  
\end{equation}
But if we are concerned with the field in a specific spatial mode $\phi(\bm{r})$, corresponding to the 
incident or reflected field mode, or defined, e.g., by the photon collection optics or a detector, 
it is easier to calculate the total photon rate in that mode \cite{Manzoni12018,Cardoner2021,Pedersen2022,Zhang2022,Kurko2021}. 
Let the field in the selected mode $\phi(\bm{r})$ with polarization $\hat{\bm{e}}$ 
be $\bm{E}(\bm{r}) = \hat{\bm{e}} \phi(\bm{r}) \hat{a}$. 
Then, for the photon rate (number of photons per unit time) emitted in that mode, 
$\expv{\hat{\alpha}^{\dagger} \hat{\alpha}} = \frac{c}{L} \expv{\hat{a}^{\dagger} \hat{a}}$, we have 
\begin{equation}
\hat{\alpha} = \hat{\alpha}_{\mathrm{in}} + \frac{i \hat{\bm{e}}^* \cdot \bm{\wp}_{ge}}{\hbar} \sqrt{ \frac{\hbar \omega} {2 \eps_0 F c} } 
\sum_j \phi^*(\bm{r}_j) \hat{\sigma}^j_{ge} , \label{eq:alpha}
\end{equation}
where $F = \int d \bm{r}^2_{\perp} |\phi(\bm{r})|^2$ and $L$ is the quantization length. 
To be specific, consider the forward Gaussian mode
\begin{equation}
\phi(\bm{r}) =  \sqrt{ \frac{\hbar \omega} {2 \eps_0 AL}}  
\frac{\zeta}{q^*(z)} \exp \left[ i k \left( z + \frac{x^2 + y^2}{2q^*(z)} \right) \right] , \label{eq:fGm}
\end{equation}
where $A = \pi w_0^2/2$ is the cross section, $w_0$ is the beam waist at the focus $z=0$, 
$\zeta = k w_0^2/2$ is the Rayleigh length,  and $q(z) = z + i \zeta$ is the complex beam parameter.
Now $F = \int d \bm{r}^2_{\perp} |\phi(\bm{r})|^2 = \frac{\hbar \omega} {2 \eps_0 L}$ and 
Eq.~(\ref{eq:alpha}) can be cast in a more intuitive form
\begin{equation}
\hat{\alpha} = \hat{\alpha}_{\mathrm{in}} + \frac{i}{ \sqrt{ \expv{\hat{\alpha}_{\mathrm{in}}^{\dagger} \hat{\alpha}_{\mathrm{in}} } } }  
\sum_j \Omega^*(\bm{r}_j) \hat{\sigma}^j_{ge} ,
\end{equation}
where $\Omega(\bm{r}) = \frac{1}{\hbar} \hat{\bm{e}} \cdot \bm{\wp}_{ge} \phi(\bm{r}) \sqrt{n}$, with 
$n = \frac{L}{c} \expv{\hat{\alpha}_{\mathrm{in}}^{\dagger} \hat{\alpha}_{\mathrm{in}}}$,  
is the Rabi frequency of the incident field at atomic position $\bm{r}$.  

Returning back to the probe field in the Gaussian mode focused at, and normal to, the atomic array,
for the transmitted field we have
\begin{equation}
\hat{\alpha}_{\mathrm{T}} = \hat{\alpha}_{p} + \frac{i}{ \sqrt{ \expv{\hat{\alpha}_p^{\dagger} \hat{\alpha}_p } } }  
\sum_j \Omega_p^*(\bm{r}_j) \hat{\sigma}^j_{ge} , \label{eq:hatalphaT}
\end{equation}
where $\Omega_p(\bm{r}) = \frac{\wp_{eg}}{\hbar} \phi(\bm{r}) \sqrt{n_p}$ with $n_p = \frac{L}{c} \expv{\hat{\alpha}_p^{\dagger} \hat{\alpha}_p }$, 
while for the field reflected in the backward propagating Gaussian mode $\phi^*(\bm{r})$, we have 
\begin{equation}
\hat{\alpha}_{\mathrm{R}} = \frac{i}{ \sqrt{ \expv{\hat{\alpha}_p^{\dagger} \hat{\alpha}_p } } }  
\sum_j \bar{\Omega}_p^*(\bm{r}_j) \hat{\sigma}^j_{ge} , \label{eq:hatalphaR}
\end{equation}
where $\bar{\Omega}_p(\bm{r}) = \frac{\wp_{eg}}{\hbar} \phi^*(\bm{r}) \sqrt{n_p}$ while the incident field in that mode is vanishing. 
Note that the field radiated by the atoms in the forward and backward directions is symmetric if all the atoms are at $z=0$, 
otherwise $\bar{\Omega}_p(\bm{r}) \neq \Omega_p(\bm{r})$ and this symmetry is broken. 

\subsection{Weak incident probe field}

We now consider a weak (coherent) probe pulse with the Gaussian temporal envelope 
$\alpha_p = \sqrt{\expv{\hat{a}_p^{\dagger} \hat{a}_p}} \frac{1}{ \sqrt{\sqrt{2 \pi} \tau}} e^{-(t/2 \tau)^2}$ 
of duration $\tau = L/c$ normalized as $\int |\alpha_p|^2 dt = \bar{n}_p = 1$, i.e., the pulse contains on average one photon. 
The field is $\sigma_{+}$ polarized in a Gaussian spatial mode (\ref{eq:fGm}) normal incident onto the atomic array. 
The atoms are assumed $^{87}$Rb, with the resonant transition $\ket{e} \to \ket{g}$ having wavelength $\lambda_e = 780\:$nm, 
free-space decay rate $\Gamma_e = 2\pi \times 6\:$MHz, and the transition dipole moment along 
$\hat{\bm{\wp}} = \frac{\hat{\bm{x}} + i \hat{\bm{y}}}{\sqrt{2}}$ (closed $\Delta M = 1$ transition).
The square 2D atomic lattice has a period $s = 0.68 \lambda_e = 532\:$nm \cite{Rui2020,Srakaew2022} and is sufficiently
larger than the probe beam waist $w_0$ at the position of the atomic sheet. 

We simulate the dynamics of the system using the stochastic quantum trajectories approach. 
We can expand the collective atomic wavefunction as 
$\ket{\Psi} = a \ket{G} + \sum_j b_j e^{-i \omega_e t} \ket{e_j} + \sum_{i<j} b^{(2)}_{ij} e^{-i 2 \omega_e t} \ket{e_i,e_j} + \ldots$,  
where $\ket{G} \equiv \ket{g_1,g_2,\ldots,g_N}$ is the collective ground state,  
$\ket{e_j} \equiv \ket{g_1,g_2,\ldots,e_j, \ldots, g_N}$ are the single excitation states,
$\ket{e_i, e_j} \equiv \ket{g_1, \ldots, e_i, \ldots, g_l, \ldots,e_j, \ldots, g_N}$ are the double excitation states, etc. 
According to the Schr\"odinder equation, the amplitudes $a,b_j,b^{(2)}_{ij},\ldots$ evolve via
\begin{subequations}
\label{eqs:atamplsge}
\begin{eqnarray}
\partial_t a &=& i \sum_j \Omega_p^*(\bm{r}_j) b_j e^{i\Delta_p t} , \\
\partial_t b_j &=& i \Omega_p(\bm{r}_j) a e^{-i\Delta_p t} + i \sum_{i \neq j} \Omega_p^* (\bm{r}_i) b^{(2)}_{ij} e^{i2\Delta_p t}
\nonumber \\ & &
- \frac{1}{2} \Gamma_e b_j - \sum_{j' \neq j} (\Gamma_{jj'} - i V_{jj'}) b_{j'} , \qquad \\
\partial_t b^{(2)}_{ij} &=& i \Omega_p(\bm{r}_i) b_j e^{-i 2 \Delta_p t} +  i \Omega_p(\bm{r}_j) b_i e^{-i2\Delta_p t} 
\nonumber \\ & &
- \Gamma_e b^{(2)}_{ij} - \sum_{j' \neq i,j} (\Gamma_{jj'} - i V_{jj'}) b^{(2)}_{ij'} 
\nonumber \\ & &
- \sum_{j' \neq i,j} (\Gamma_{ij'} - i V_{ij'}) b^{(2)}_{j'j} , \\
 \ldots & &  , \nonumber
\end{eqnarray}
\end{subequations}
interrupted by quantum jumps which project the system onto the state with one less excitation. 
The normalized atomic amplitudes at any time are $\bar{a}(t) = a(t)/|| \Psi(t) ||$,
$\bar{b}_j(t) = b_j(t)/|| \Psi(t) ||$, $\bar{b}^{(2)}_{ij}(t) =  b^{(2)}_{ij}/|| \Psi(t) ||$, etc.,
while the amplitudes of the transmitted and reflected fields are
\begin{subequations}
\label{eq:alphaTR}
\begin{eqnarray}
\alpha_{\mathrm{T}} &=& \alpha_{p} + \frac{i}{|\alpha_p|}  
\sum_j \Omega_p^*(\bm{r}_j) \expv{\tilde{\sigma}^j_{ge}} , \label{eq:alphaT} \\
\alpha_{\mathrm{R}} &=& \frac{i}{| \alpha_p|}  
\sum_j \bar{\Omega}_p^*(\bm{r}_j) \expv{\tilde{\sigma}^j_{ge}} , \label{eq:alphaR}
\end{eqnarray}
\end{subequations}
where the slowly-varying atomic polarizations are given by
\begin{equation}
\expv{\tilde{\sigma}^j_{ge}} = \bar{a}^*\tilde{b}_j + \sum_{i\neq j} \tilde{b}_i^{^*} \tilde{b}^{(2)}_{ij} + \ldots . \label{eq:expvsigj}
\end{equation}
with $\tilde{b}_j = \bar{b}_j e^{i \Delta_p t}$, $\tilde{b}^{(2)}_{ij} = \bar{b}^{(2)}_{ij} e^{i \Delta_p t}$, etc.
The total transmission and reflection probabilities $p_{\mathrm{T,R}}$ for the probe pulse are obtained via the integration 
$p_{\mathrm{T,R}} = \int dt |\alpha_{\mathrm{T,R}}|^2$, while the scattering probability is $p_{\mathrm{S}} = 1 - p_{\mathrm{T}} - p_{\mathrm{R}}$.

We first perform simulations of the atomic amplitude Eqs.~(\ref{eqs:atamplsge}) for a small system of $N \gtrsim 100$ atoms 
without quantum jumps, i.e., assuming conditional no-jump dynamics \cite{PLDP2007,QMCWF}.
We verify that, for a weak incoming pulse of duration $\tau \simeq 2\:\mu$s containing on average $\bar{n}_p = 1$ photon 
[the probabilities $p_n$ of $n$-photon states are $p_{n=0,1,2,3,\ldots} = (1,1,1/2,1/6, \ldots) \times e^{-1}$], 
the total probability of double excitations $P_{2e} = \sum_{i<j} |\bar{b}^{(2)}_{ij}|^2$ remains small at all times, 
with the peak value $P_{2e} \simeq 10^{-4}$, while the total probability of single excitations 
$P_{1e} = \sum_{j} |\bar{b}_{j}|^2$ attains peak values $P_{1e} \simeq 2 \times 10^{-2}$. 
This means that double and multiple excitation contribute little ($\sim 0.5\%$) to the atomic polarization (\ref{eq:expvsigj}). 
We can therefore neglect the amplitudes $b^{(2)}_{ij}$ of multiple atomic excitations which makes the Monte Carlo simulations of
Eqs.~(\ref{eqs:atamplsge}) particularly efficient and tractable even for large $N \sim 500$. Then any quantum jump projects 
the system onto the collective ground state $\ket{G}$, since we have at most a single collective excitation, 
and we do not need to determine the various collective decay channels and their jump probabilities, as described above.

In Fig.~\ref{fig:scheme0}(c) we show the transmission, reflection and scattering spectrum for the probe field. 
We observe a collective resonance at $\Delta_p = \Delta \simeq 0.17 \Gamma_e$ and the position $\Delta$ and width $\Gamma$
of that resonance depend on the lattice constant; more precisely, they depend on the surface density of the atoms, 
and are independent of the lattice geometry (square, triangular) \cite{Shahmoon2017}, and with increasing the density 
(decreasing $s$) $\Delta$ shifts to the blue side (larger $\Delta > 0$) and the resonance broadens, 
$\Gamma = \frac{3}{4\pi} \frac{\lambda_e^2}{a^2} \Gamma_e$.      

Next we observe that on resonance the transmission vanishes and the reflection probability attains a large value $p_{\mathrm{R}} \gtrsim 0.98$
which is however slightly smaller than 1 since we truncated the multiple atomic excitations. 
We have verified that by decreasing the pulse amplitude $\alpha_p$, by decreasing the mean photon number 
$\bar{n}_p=\expv{\hat{a}_p^{\dagger} \hat{a}_p} \ll 1$ or increasing the pulse duration $\tau$,
the reflection probability increases, approaching $p_{\mathrm{R}} \to 1$ as predicted by the linear response and static atomic polarizability
theory for infinite lattice radiated by a weak cw plane wave field \cite{Shahmoon2017}.

Finally, we simulate the dynamics of the system in the presence of position disorder of the atoms, see the inset of Fig.~\ref{fig:scheme0}(c). 
We observe that just a few percent of position disorder with respect to the lattice constant $s$, or wavelength $\lambda_e$, 
already significantly degrade the performance of the atomic mirror: 
for standard deviation of the disorder $\sigma_{xy,z} \gtrsim 0.025s \simeq 20\;$nm the reflection probability is reduced to $p_{\mathrm{R}} \lesssim 0.65$
consistent with \cite{Rui2020,Srakaew2022}. For still larger disorder, we observe that the atomic position disorder $\sigma_{z}$
along the field propagation axis $z$ (surface roughness) leads to more scattering of the light than a similar position disorder 
$\sigma_{xy}$ in the normal plane $xy$.


\section{Hybrid system involving multilevel atoms}

Having reviewed the transmission and reflection of a weak probe pulse from a two-dimensional array of two-level atoms, 
we next consider a hybrid quantum system that integrates such atomic arrays with superconducting and optical elements.   

\begin{figure}[t]
\centering
\includegraphics[width=1.0\linewidth]{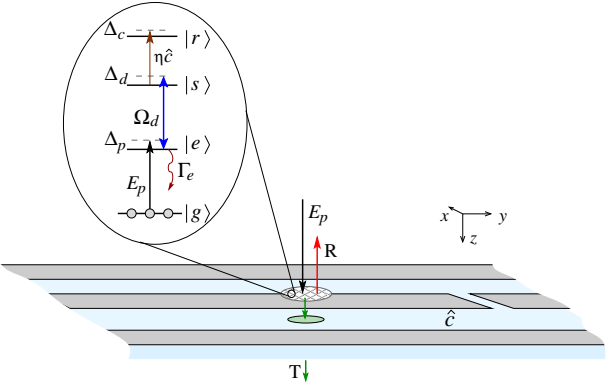}
\caption{Schematics of the hybrid system:  
The two-dimensional array of atoms is positioned near a superconducting microwave coplanar waveguide resonator 
while the incoming probe field $E_p$ is reflected or transmitted into an optical waveguide whose 
collection lens is placed under the atomic array in the gap between the superconducting elements of the resonator. 
The inset shows the atomic level scheme:
The atoms initially in the ground state $\ket{g}$ interact with the probe field 
on the transition $\ket{g} \to \ket{e}$ with detuning $\Delta_p$, 
the electronically excited state $\ket{e}$ is coupled to a Rydberg state $\ket{s}$ 
by a classical driving field with Rabi frequency $\Omega_d$ and detuning $\Delta_d$,
while the Rydberg transition $\ket{s} \to \ket{r}$ is strongly coupled to the
microwave cavity mode $\hat{c}$ with strength $\eta$ and detuning $\Delta_c$.
The presence or absence of a microwave cavity photon changes the transmission 
and reflection of the atomic array.}
\label{fig:SCpOL}
\end{figure}

Our aim is to extend the functionalities of superconducting atom chips which contain microwave resonators 
to mediate coupling between superconducting qubits \cite{Clerk2020,Blais2004,Schoelkopf2008,Clerk2020}  
while simultaneously interacting with cold trapped atoms \cite{Hattermann2017,Kaiser2022}.
We envisage a setup sketched in Fig.~\ref{fig:SCpOL}: The atomic array is positioned in the vicinity of a coplanar waveguide resonator 
that also incorporates optical elements for the collection of the transmitted probe field into an optical waveguide.
A strong classical field with wavevector $\bm{k}_d$ and frequency $\omega_d$ drives the atomic transition 
from the electronically excited state $\ket{e}$ to the Rydberg state $\ket{s}$ 
with Rabi frequency $\Omega_d$ and detuning $\Delta_d = \omega_d - \omega_{se}$. 
The resonator microwave field mode $\hat{c}$ with frequency $\omega_c$ strongly couples to the atoms 
on a dipole-allowed transition between the Rydberg states $\ket{s}$ and $\ket{r}$ 
with detuning $\Delta_c = \omega_c-\omega_{rs}$ and strength $\eta$ (vacuum Rabi frequency).
The coupling strength $\eta(\bm{r}) =(\wp_{rs}/\hbar) \veps_{c} \phi_c(\bm{r})$ is proportional to the dipole moment $\wp_{rs}$ of the atomic transition, 
the field per photon $\veps_c =\sqrt{\hbar \omega_c/\eps_0 V_c}$ in the cavity with effective volume $V_c$, and the cavity mode 
function $\phi_c(\bm{r})$ at the position $\bm{r}$ of the atoms. 
We assume the parameters similar to those in \cite{Hattermann2017,Kaiser2022,Petrosyan2019}:
With the strip-line length $l=10.5\:$mm and the gap width $d=10\:\mu$m to the grounded electrodes, 
the effective cavity volume is $V_c \simeq 2 \pi d^2 l$ yielding the field per photon $\veps_c = \sqrt{\hbar \om_c/\eps_0 V_c } \simeq 0.37\:$V/m 
for the full-wavelength cavity mode of frequency $\om_c/2\pi = c/l\sqrt{\eps_r} \simeq 12\:$GHz ($\eps_r \simeq 5.6$). 
The atoms are placed at the antinode of the standing-wave cavity field which falls off evanescently with the distance from the chip surface, 
$\phi_c(z) \simeq e^{-|z|/d}$. We choose the Rydberg states $\ket{i} = \ket{68P_{3/2},m_J=1/2}$ and $\ket{s} = \ket{69S_{1/2},m_J=1/2}$ 
of Rb having the transition frequency $\om_{rs}/2\pi \simeq 12\:$GHz and dipole moment $\wp_{si} \simeq 2000a_0 e$ \cite{RydAtoms}. 
Then for the atoms at $z=10-15\:\mu$m from the chip surface the coupling strength to the cavity mode is $\eta(z)/2\pi \simeq 2-4\:$MHz.

The total Hamiltonian $H = H_{\mathrm{af}} + H_{\mathrm{RDDI}} + H_{\mathrm{d}} + H_{\mathrm{c}}$ acquires now two new terms,
\begin{equation}
H_{\mathrm{d}} = \hbar \sum_j^N \{ \omega_s \hat{\sigma}_{ss}^{j} - [\Omega_d e^{i\bm{k}_d \cdot \bm{r}_i} \hat{\sigma}_{se}^{j} + \mathrm{H. c.}] \},
\end{equation}
due to the coupling with the spatially uniform driving field, and
\begin{equation}
H_{\mathrm{c}} = \hbar \sum_j^N \{ \omega_r \hat{\sigma}_{rr}^{j} - [\eta(\bm{r}_j) \hat{c} \, \hat{\sigma}_{rs}^{j} + \mathrm{H. c.}] \},
\end{equation}
due to the coupling to the microwave cavity.
Consistency with the discussion above, we can expand the collective atomic wavefunction in the basis of at most single excitations as
$\ket{\Psi} = a \ket{G} + \sum_j b_j e^{-i \omega_e t} \ket{e_j} + \sum_{j} c_{j} e^{-i \omega_s t} \ket{s_j} +  \sum_{j} d_{j} e^{-i \omega_r t} \ket{r_j}$,  
where $\ket{s_j} \equiv \ket{g_1,g_2,\ldots,s_j, \ldots, g_N}$ and $\ket{r_j} \equiv \ket{g_1,g_2,\ldots,r_j, \ldots, g_N}$ are the single 
Rydberg excitation states. Then, according to the Schr\"odinder equation, the amplitudes $a,b_j,c_j,d_j$ evolve via
\begin{subequations}
\label{eqs:atamplsgesr}
\begin{eqnarray}
\partial_t a &=& i \sum_j \Omega_p^*(\bm{r}_j) b_j e^{i\Delta_p t} , \\
\partial_t b_j &=& i \Omega_p(\bm{r}_j) a e^{-i\Delta_p t} + i \Omega_d^* e^{-i\bm{k}_d \cdot \bm{r}_i} c_{j} e^{i\Delta_d t}
\nonumber \\ & &
- \frac{1}{2} \Gamma_e b_j - \sum_{j' \neq j} (\Gamma_{jj'} - i V_{jj'}) b_{j'} , \qquad \\
\partial_t c_j &=& i \Omega_d e^{i\bm{k}_d \cdot \bm{r}_i} b_{j} e^{-i\Delta_d t} + i \Omega_c^* d_j e^{i\Delta_c t}  - \frac{1}{2} \Gamma_s c_j , \qquad \\
\partial_t d_j &=& i \Omega_c c_j e^{-i\Delta_c t}  - \frac{1}{2} \Gamma_r d_j ,
\end{eqnarray}
\end{subequations}
where we included the small decay rates of the Rydberg states $\Gamma_{s,r} \ll \Gamma_e$, while 
$\Omega_c = \eta \sqrt{n_c}$ is the Rabi frequency of the cavity field containing $n_c = \expv{c^{\dagger} c}$ microwave photons
and we assume that the cavity mode function $\phi_c(\bm{r})$, and thereby $\eta$, varies little across the atomic array 
which is parallel to the chip surface with the dimension much smaller than the wavelength of the microwave radiation $\la_c \sim l$.

\begin{figure}[t]
\centering
\includegraphics[width=1.0\linewidth]{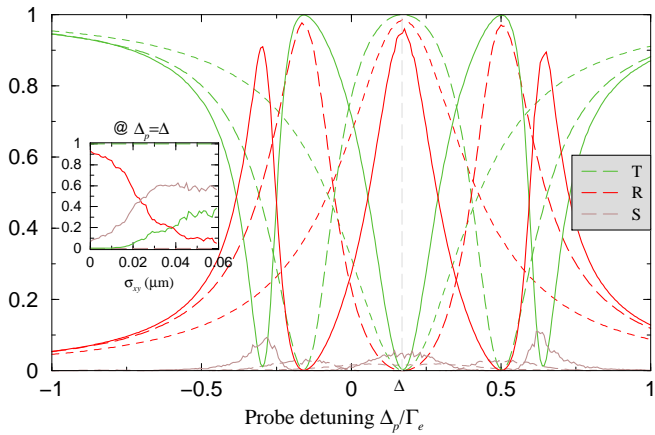}
\caption{Transmission (T, green), reflection (R, red) and scattering (S, brown) spectra 
of the two-dimensional array of four-level atoms for an incoming coherent probe pulse ($\bar{n}_p = 1$) of duration $\tau = 2\:\mu$s 
focused at the atomic layer in the $xy$ plane. 
The driving field has the Rabi frequency $\Omega_d= 2\pi \times 2.0\;$MHz and detuning $\Delta_d = -0.172\Gamma_e$ to provide
a two-photon resonance for the probe field at the collective resonance frequency $\Delta_p = \Delta=0.172\Gamma_e$.
The cavity mode is assumed resonant, $\Delta_c=0$, and couples to the Rydberg transition with strength $\eta = 2\pi \times 2.0\;$MHz.
The decay rates of the Rydberg states are $\Gamma_{s,r} = 10^{-3} \Gamma_e$ with the other parameters as in Fig.~\ref{fig:scheme0}.
The cavity mode is empty, $n_c=0$ and $\Omega_c =0$, (long dashed lines); or contains one photon, $n_c=1$ and $\Omega_c = \eta$, (solid lines). 
For reference, we also show the response of two level atoms as in Fig.~\ref{fig:scheme0}(c) (thin dashed lines).  
Inset shows the transmission, reflection and scattering of the probe field at the collective resonance frequency $\Delta_p = \Delta$
vs the atomic position uncertainly $\sigma_x = \sigma_y$ while $\sigma_z = 0.01\:\mu$m. 
The graph are obtained from Monte Carlo simulations of Eqs.~(\ref{eqs:atamplsgesr}) in conjunction with Eqs. (\ref{eq:alphaTR}) and (\ref{eq:expvsigj}).}
\label{fig:TRAEIT}
\end{figure}

\paragraph{Resonant drive and cavity fields.} 
In Fig.~\ref{fig:TRAEIT} we show the transmission, reflection and scattering spectra of the four-level atomic medium for the incoming probe pulse.
When the microwave cavity field mode is empty, $n_c=0$ and $\Omega_c=0$, the coherent drive on the atomic transition $\ket{e} \to \ket{s}$ 
to the long-lived Rydberg state $\ket{s}$ results in the Autler-Townes splitting of the atomic resonance by $\pm \Omega_d$
(assuming $|\Delta_d| \ll \Gamma_e$) resulting in electromagnetically induced transparency (EIT) for a resonant probe \cite{EITrev2005}.
Essentially the probe field with frequency within the EIT window feels no atoms.
We tune the detuning of the driving field to $\Delta_d = - \Delta$ to be at the two-photon resonance on the transition
$\ket{g} \to \ket{s}$ and obtain for the probe field at the collective resonance frequency $\Delta_p = \Delta$
perfect transmission, $p_{\mathrm{T}} \simeq 1$, $p_{\mathrm{R,S}} \simeq 0$, even through a disordered atomic array.    
When, however, the resonant cavity mode is populated by one or more photons, $n_c \geq 1$ and $\Omega_c = \eta \sqrt{n_c}$, 
it spits the EIT resonance by $\pm \Omega_c$ and the perfect transmission of the probe pulse turns 
to a strong reflection at $\Delta_p = \Delta$, as for a two-level atomic medium. 
Now, again, the reflection resonance is sensitive to atomic position disorder, and already for
$\sigma_{x,y,z} = 10\:$nm we obtain $p_{\mathrm{T}} \simeq 0$, $p_{\mathrm{R}} \simeq 0.85$, and $p_{\mathrm{S}} \simeq 0.15$.

\begin{figure}[t]
\centering
\includegraphics[width=1.0\linewidth]{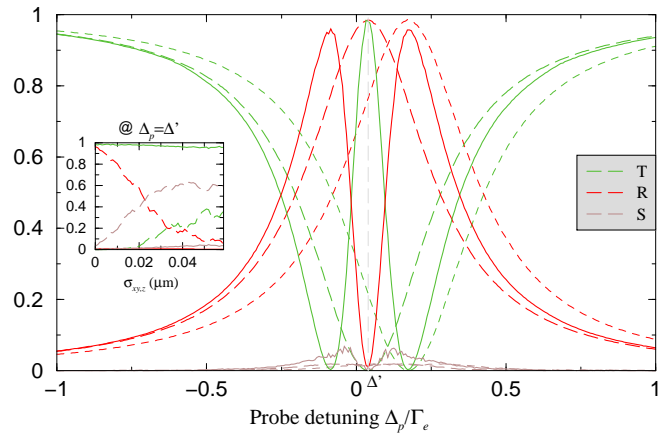}
\caption{Transmission (T, green), reflection (R, red) and scattering (S, brown) spectra of the array of atoms for an incoming probe pulse. 
The driving field has the Rabi frequency $\Omega_d= 2\pi \times 4.0\;$MHz and detuning $\Delta_d = - 2\pi \times 20.0\;$MHz
and the cavity mode has the coupling strength $\eta= 2\pi \times 4.0\;$MHz and detuning $\Delta_c = - \Delta_d - \Delta \simeq 2\pi \times 19.0\;$MHz.
All the other parameters are as in Fig~\ref{fig:TRAEIT}.
When the cavity mode is empty, $n_c=0$ and $\Omega_c =0$, (long dashed lines), the probe field undergoes nearly perfect reflection from the
atomic array at the AC Stark shifted collective resonance frequency $\Delta_p = \Delta' \equiv \Delta + S_e$. 
When the cavity mode contains a photon, $n_c=1$ and $\Omega_c = \eta$, (solid lines), together with 
the driving field it results in EIT for the probe field in the vicinity of  $\Delta_p = \Delta'$. 
For reference, we also show the response of two level atoms as in Fig.~\ref{fig:scheme0}(c) (thin dashed lines).  
Inset shows the transmission, reflection and scattering of the probe field at frequency $\Delta_p = \Delta'$
vs the atomic position uncertainly $\sigma_x = \sigma_y$ while $\sigma_z = 0.01\:\mu$m.}
\label{fig:TRAEITD}
\end{figure}

\paragraph{Non-resonant drive and cavity fields.}
We next consider a slightly different scheme: We assume that both the driving field and the cavity mode are largely
detuned from the corresponding atomic transitions, $|\Delta_d| \gg \Omega_d$ and $|\Delta_c| \gg \eta$, but that their detunings satisfy 
$\Delta_d + \Delta_c = - \Delta$ corresponding to a two-photon resonance on the transition $\ket{e} \to \ket{r}$ with small detuning $-\Delta$.
Adiabatically eliminating the non-resonant intermediate state $\ket{s}$ we then obtain the AC Stark shifts 
$S_e = |\Omega_d|^2/\Delta_d$ of level $\ket{e}$ and $S_r = - |\Omega_c|^2/\Delta_c$ of level $\ket{r}$.  
When the cavity mode is empty, $n_c =0$ and $\Omega_c=0$, the detuned driving field does not result in EIT but only shifts 
the collective reflection resonance of the two-level atoms to $\Delta_p = \Delta' \equiv \Delta + S_e$ 
(the collective resonance is still at detuning $\Delta$ from the Stark-shifted atomic transition $\ket{g} \to \ket{e}$), 
as seen in Fig.~\ref{fig:TRAEITD}. The sensitivity of the reflection probability to the atomic position disorder is similar to that 
for two-level atoms. 
But when the cavity mode contains a photon, $n_c = 1$ and $\Omega_c = \eta$, together with the driving field
it results in the EIT for the probe field via the two-photon driving of the $\ket{e} \to \ket{r}$ transition with 
the effective Rabi frequency $\Omega^{(2)} = \Omega_c \Omega_d/\Delta_d$ and detuning $-\Delta + S_e - S_r$. 
Assuming $S_e \simeq S_r$, we then have a three-photon resonance $\ket{g} \to \ket{r}$ for the probe field with 
the frequency $\Delta_p = \Delta'$. Now the probe field with the frequency within the EIT window is transmitted with 
nearly unit probability, $p_{\mathrm{T}} \simeq 1$, $p_{\mathrm{R,S}} \simeq 0$, and is insensitive to the atomic position disorder. 
Of course for larger cavity photon number, $n_c >1$, the AC Stark shift of level $\ket{r}$ will be different, $S_r = - |\eta|^2n_c/\Delta_c$,
which in principle can be compensated by adjusting the frequency of the driving field to satisfy the three-photon resonance conditions 
and attain EIT. But if our aim is to realize a switch for the probe field controlled by the presence or absence of a microwave cavity
photon, this scheme should be applied for $n_c = 0$ or 1. 

To summarize, in scheme (\textit{a}) the optical probe pulse is transmitted through the atomic array when the microwave cavity is empty and 
is reflected when the cavity has one or more photons, and vice versa in scheme~(\textit{b}). For a perfectly ordered array, the reflection
probability in scheme~(\textit{b}) is closer to unity than in scheme~(\textit{a}) and in both schemes it rapidly decreases with increasing
the position uncertainties $\sigma_{x,y}$ in the plane and even more so with the position uncertainly $\sigma_{z}$ in the field propagation direction.
In both schemes, the transmission provided by the EIT is nearly perfect and is insensitive to the position uncertainly of the atoms in the array. 
In all cases, good performance of the switch requires that the bandwidth of the probe pulse of duration $\tau$ be small compared to the width 
of the reflection resonance, strength of the atom-cavity coupling and the bandwidth of the EIT \cite{EITrev2005}: 
$2\pi/\tau < |\Omega_{d,c}|^2/\Gamma_e$ for scheme (\textit{a}), and $2\pi/\tau < \Gamma, |\Omega^{(2)}|^2/\Gamma_e$ for scheme (\textit{b}).

\section{Conclusions}

To conclude, we have studied a hybrid quantum system composed of a two-dimensional array of atoms near an integrated superconducting chip
containing microwave coplanar waveguide resonator and optical elements for strong coherent coupling of optical and microwave photons.
Our scheme allows to implement high-fidelity transmission or reflection of optical photons by the atomic array controlled by the presence 
or absence of microwave photons in the cavity that couples strongly to the atoms on a resonant Rydberg transition with strong electric dipole 
moment. Quantum interfaces between microwave and optical fields are interesting and important for various quantum technology applications,
including realization of quantum communications and quantum Internet between distant quantum computers based on superconducting circuits
using optical photons propagating over long distances in optical waveguides with little loss.

\section*{Acknowledgments}
We thank Ephraim Shahmoon, Michael Fleischhauer, and Christian Gro{\ss} for enlightening discussions.
This work was supported by the EU QuantERA projects MOCA and PACE-IN (GSRT grant No. T11EPA4-00015).
D.P. was also supported by EU HORIZON-RIA project EuRyQa.
D.P. and J.F. were also supported by DFG (SPP 1929 GiRyd and FOR-5413). 
G. K. was also supported by EU FET Open project PATHOS and DFG (FOR-7274).


\end{document}